\documentclass[preprint2]{aastex}
\usepackage{array}
\usepackage{multirow}

\begin{document}
\title{Metallicity Gradients of Thick Disk Dwarf Stars}
\author{Kenneth Carrell, Yuqin Chen and Gang Zhao}
\affil{Key Laboratory of Optical Astronomy, National Astronomical Observatories,
Chinese Academy of Sciences, Beijing 100012, China}
\email{carrell@nao.cas.cn}

\begin{abstract}
We examine the metallicity distribution of the Galactic thick disk using 
F, G, and K dwarf stars selected from the Sloan Digital Sky Survey, 
Data Release 8. Using the large sample of dwarf stars with proper motions 
and spectroscopically determined stellar parameters, metallicity 
gradients in the radial direction for various heights above the Galactic
plane and in the vertical direction for various radial distances from 
the Galaxy center have been found. In particular, we find a 
vertical metallicity gradient of 
--0.113~$\pm$~0.010 (--0.125~$\pm$~0.008) dex~kpc$^{-1}$ using an 
isochrone (photometric) distance determination in the range 
1~$<|Z|<$~3~kpc, which is the vertical height range most consistent 
with the thick disk of our Galaxy. In the radial direction, we find 
metallicity gradients 
between +0.02 and +0.03 dex~kpc$^{-1}$ for bins in the vertical 
direction between 1~$<|Z|<$~3 kpc.  Both of these 
results agree with similar values determined from other populations of 
stars, but this is the first time a radial metallicity gradient for the 
thick disk has been found at these vertical heights. We are also able 
to separate thin and thick disk stars based on kinematic and spatial 
probabilities in the vertical height range where there is significant 
overlap of these two populations. This should aid further studies of 
the metallicity gradients of the disk for vertical heights lower than 
those studied here but above the solar neighborhood. Metallicity 
gradients in the thin and thick disks are important probes into 
possible formation scenarios for our Galaxy and a consistent picture 
is beginning to emerge from results using large spectroscopic surveys, 
such as the ones presented here.
\end{abstract}

\section{Introduction\label{sec:intro}}
The thick disk of our Galaxy has been extensively studied (especially
in the solar neighborhood) since it was discovered by \citet{gil83}.
Photometric and spectroscopic studies have shown that it is a distinct
component in a number of ways. \citet{jur08} find that the stellar 
number density distribution of dwarf stars in the Galactic disk found 
using Sloan Digital Sky Survey (SDSS) data and a photometric parallax 
method is well fit with a two-component exponential. The thick disk 
component is also known to have both a rotational lag and a larger 
velocity dispersion than the thin disk \citep{chi00,sou03} giving 
it a distinct kinematic signature. It has further been shown to be 
the dominant component for stars with vertical distances of 
1~$<|Z|<$~3~kpc from the Galactic plane \citep{all06}.

The metallicity distribution of the thick disk is still an unresolved 
issue, however. Using the metallicity and kinematics of SDSS stars, 
\citet{ive08} are able to reproduce the results of previous works, 
but also caution that there are problems with describing their 
data using a two-component exponential. Most importantly, they 
suggest that the metallicity and velocity distributions can be 
used to describe a single disk component instead of two separate 
ones. This would suggest a complex nature to a single disk 
component and not two distinct disk components with different 
formations and evolutions. More recently, \citet{bovy12} also find 
that the Galactic disk can be represented by a single exponential 
in the radial and vertical directions that is smoothly varying as 
a function of age (by proxy using abundances).

This has naturally lead to several competing
theories as to the formation mechanism of the thick disk and how it
relates to the thin disk. For example, the thick disk could be mainly
composed of accreted stars from previous mergers \citep{aba03}, composed
of stars formed from accreted gas from previous mergers \citep{bro04},
or from the internal mixing of stars and gas within the Galaxy 
\citep{sch09}. These are just a few of the various internal and external 
sources from which the thick disk of our Galaxy could arise. Obviously, 
a combination of some or all of these scenarios could also give rise
to our current Galactic hierarchy. Certain well-defined properties
of the thick disk should be able to distinguish between these possible
scenarios and provide a clearer picture of the formation and evolution
of our Galactic disk.

In the radial direction, metallicity gradients vary from 
$\sim$0~dex~kpc$^{-1}$ \citep{all06} to 
+0.028~$\pm$~0.036 dex~kpc$^{-1}$ \citep{nor04}
for stars likely to be thick disk members. We note that the former
uses spectroscopically derived stellar parameters of F- and G-type
stars observed by SDSS in the range
1~$<|Z|<$~3~kpc while the latter uses photometrically derived
stellar parameters of nearby F- and G-type stars. Recent results from
the Radial Velocity Experiment \citep[RAVE,][]{rave} fall within these
two values \citep{ruc11,cos12}.

Vertical metallicity gradients of the thick disk show more discrepancy.
\citet{all06} found no variation of metallicity as a function of
vertical height. Using SDSS Red Horizontal Branch stars, \citet{chen11}
found a gradient as high as --0.22~$\pm$~0.07 dex~kpc$^{-1}$ but this 
result most likely suffered contamination from other Galactic sources. 
Again, the RAVE results fall somewhere in the middle \citep{kor11,ruc11}
as do the results of \citet{katz11} who observed stars along two
different lines of sight.

Our aim in this work is to examine both radial and vertical metallicity
gradients of stars associated with the thick disk of our
Galaxy. We use the large sample of spectroscopically observed stars
from the Sloan Extension for Galactic Understanding and Exploration
(SEGUE) made available in SDSS Data Release 8 (DR8). Selecting
dwarf stars in this analysis is important since it is the dominant
component of the disk of our Galaxy. Furthermore, these stars have
changed very little since their formation, making them an excellent
probe of the condition of our Galaxy during the formation of its disk.
We employ various techniques used by previous analyses in order to
more easily compare results from different sets of data and using
different selection criteria. The large sample of stars with reliable
parameter determinations facilitates this comparison and allows us
to give a more comprehensive view of the metallicity distribution
of stars up to several kiloparsecs above or below the Galactic plane.

\section{Data\label{sec:data}}
\begin{figure}[tb]
\plotone{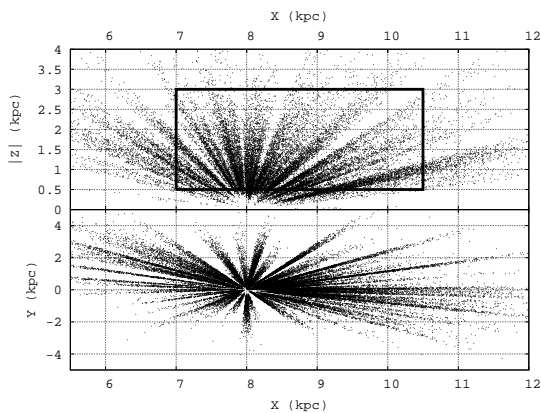}
\caption{Spatial distribution of F, G, and K dwarf stars selected 
from SDSS DR8.\label{fig:spatial}}
\end{figure}

Spectroscopic data from SDSS DR8 \citep{aih11} was obtained from
the \texttt{CasJobs} interface. The improvement in DR8 for high-metallicity
stars is important for this work since we expect many stars to have
abundances near or slightly below solar. We used the \textit{segue1\_target1}
flag to select F, G, and K dwarfs (integer values of --2147483136,
--2147221504, and --2147450880, respectively). We require that the stars
have clean photometry and reasonable spectroscopic parameters (wide
cuts were applied to eliminate any undetermined or extreme values
for the effective temperature, surface gravity, metallicity and radial
velocity). This leaves us with 66,481 out of the total 88,081 stars
for these three categories. Further cuts were applied to ensure a
clean sample of dwarf stars with good quality observed values and
are summarized in Table \ref{tab:cuts}. Our final sample is 43,417 
F, G, and K dwarf stars (slightly less than half the full sample)
that have well defined photometric and spectroscopic values as well
as reasonable proper motions. In Figure \ref{fig:spatial} we show
the spatial distribution of our dwarf star sample using distances 
calculated as described below.

\begin{table*}
\begin{center}
\caption{Selection cuts for our data.\label{tab:cuts}}
\begin{tabular}{c c c c}
\hline 
\textbf{Category} & \textbf{SDSS Criteria}
 & \textbf{Stars} & \textbf{Our Criteria} \\
\hline
\multirow{2}{*}{F/G } & $g<20.2$ & \multirow{2}{*}{6939}
 & $12.0<g,r,i<21.0$ mag \\
 & $0.2 <(g-r)<0.48$  &  & $g,r,i$ error $<0.05$ mag \\
\cline{1-3}
\multirow{2}{*}{G dwarf} & $14.0<r<20.2$ & \multirow{2}{*}{62784}
 & $\log g>4.4$ dex AND $\log g$ error $<0.25$ dex \\
 & $0.48<(g-r)<0.55$  &  & T$_{eff}$ error $<250$ K \\
\cline{1-3}
\multirow{2}{*}{K dwarf} & $14.5<r<19.0$ & \multirow{2}{*}{18358}
 & [Fe/H] error $<0.25$ dex \\
 & $0.55<(g-r)<0.75$ &  & radial velocity error $<10$ km s$^{-1}$ \\
 &  &  & proper motion errors $<6$ mas yr$^{-1}$ \\
\hline 
 \textbf{TOTALS} &  & 88,081 & 43,417 \\
\hline 
\end{tabular}
\end{center}
\end{table*}

\begin{figure}[tb]
\plotone{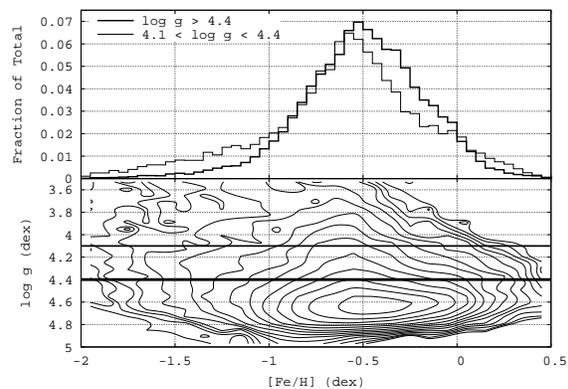}
\caption{Surface gravities and metallicities for the F, G, 
and K dwarf stars selected from SDSS DR8. The horizontal lines in the
lower plot correspond to surface gravities of 4.1 and 4.4 dex. The
upper plot shows histograms of the metallicity for two surface gravity
selections.\label{fig:loggVabun}}
\end{figure}

The most stringent cut for our dwarf sample is on the surface gravity
since we want a sample free of possible turnoff or giant stars to
aid in the reliability of the distance determination. Therefore, selecting
a cut on the surface gravity of 4.4~dex and requiring that the error
on this value be less than 0.25~dex ensures that we will not have
stars with surface gravities lower than $\sim$4.15~dex. Furthermore,
as can be seen in Figure \ref{fig:loggVabun}, there is a noticeable
change in the distribution of the metallicity as a function of 
surface gravity near $\log g\sim$~4.4~dex. The histogram in the
upper portion of Figure \ref{fig:loggVabun} clearly shows that the
metallicity distribution of stars with 4.1~$< \log g <$~4.4~dex
is wider and has a peak that is more metal poor than for stars with
$\log g >$~4.4~dex. We note that \citet{cheng12} concentrate on
the main sequence turnoff stars that we are trying to eliminate. Also,
\citet{sch11} use G and K dwarfs in their analysis, but their surface
gravity cut is $\log g>$~4.1~dex. This means that their sample
overlaps in surface gravities with this paper and with \citet{cheng12}.
The much steeper vertical metallicity gradient in \citet{sch11} could
be due to this fact as well as the modifications they apply to the
metallicities provided by the SEGUE Stellar Parameter Pipeline and
the weighting algorithm they use (both of which tend toward higher
metallicity).

Selection bias in SDSS spectroscopic samples can be extremely difficult
to determine or even estimate. Our sample comes from a relatively
simple set of selection criteria so we expect the bias to be minimal.
Essentially, targets were selected from a wide color range 
(0.2~$<(g-r)<$~0.75)
with more weight given to G type stars. Since we are not interested
in a specific population, but in the dwarf stars as a whole, the number
of each spectral type is not a primary concern. What is most essential
is that the selection criteria should not influence the expected abundances
of our sample stars, and a simple color cut should not introduce a
significant bias in this regard.

Distance determinations were made using two methods. The first method 
we used was adapted from Equation 1 of \citet{kor11},
which is based on the method found in \citet{zwi10}. This technique
uses theoretical isochrones to determine the absolute magnitude of
a star with a certain set of parameters and their errors. This is,
in essence, a Bayesian approach with the likelihood determined by
the distance from a point on the isochrone and the prior probability
determined by the mass distribution. We used \citet{gir04} isochrones
that were downloaded from the 
CMD\footnote{http://stev.oapd.inaf.it/cgi-bin/cmd} web interface. The 
grid of isochrones has ages of 2--12 Gyr in 1 Gyr increments and is 
linear in Z (logarithmic in [Fe/H]) from Z~=~0.0001 to 0.03 in 
0.0001 increments. The range in [Fe/H] of roughly 
--2.25~$<$~[Fe/H]~$<$~0.2 dex includes the expected metallicities
of the dwarf stars in our sample and the reliable limits of DR8. The
large age range is necessary if we want to accurately determine distances
for both the (younger) thin disk and (older) thick disk stars, but
since our sample is composed of only dwarf stars, this age range has
little effect because all the stars in our sample should be main sequence
stars.

Using the mass weighted method described above, the probability distribution
for each star was determined using four observables: the spectroscopically
determined effective temperature, surface gravity, and metallicity
and the photometrically determined $(g-r)_0$ color. Individual errors
on each parameter were used instead of overall errors quoted for DR8
measurements to give a better indication of the intrinsic uncertainty
of the distance determination for each star. Once the probability
distribution was calculated, the most probable absolute magnitude
for the star was found from a weighted average and the error on this
value was determined from the weighted standard deviation. Using this
absolute magnitude and the measured apparent magnitude one can find
the distance to the star and convert to the Galactic Cartesian system
(X,Y,Z) using its position on the sky. As a further quality control
of our stellar sample, we only use stars that have distance errors
from the isochrone method of less than 5\%. This eliminates less than 
0.1\% of the sample, but does help identify and remove outliers and
stars with inconsistent observations or parameter determinations.

\begin{figure}[tb]
\plotone{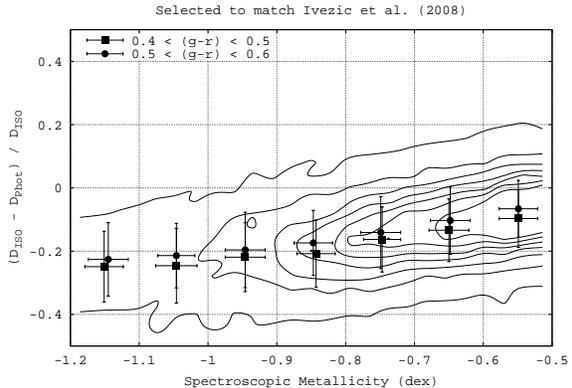}
\caption{Fractional difference between distances determined using 
the isochrone method of \citet{kor11} and the photometric method 
of \citet{ive08} as a function of metallicity for two color 
ranges.\label{fig:distdiff}}
\end{figure}

The other distance determination we used came from Equation 1 of 
\citet{ive08}. This method determines the absolute $r$ magnitude of a 
star using its $(g-i)$ color and metallicity. We use this method 
as a comparison to the isochrone method to help identify systematic 
errors in the distance determination. In Figure \ref{fig:distdiff} 
we plot the fractional difference between the distances 
determined using the isochrone and photometric methods as a function 
of metallicity for two color selections. In this plot we have used 
the same selection criteria as that used by \citet{ive08} and 
determine distances using photometric metallicities 
derived using the equations from their work. The range of 
metallicities in the plot is limited to the 
reliable range found by \citet{ive08} for their photometric 
abundance determination. We find that the distances derived 
using the method of \citet{ive08} are systematically farther 
than the isochrone method of \citet{kor11} by $\sim$10\% on the 
high metallicity end ([Fe/H]~$\sim$~--0.55~dex) and increase to 
$\sim$20\% at lower metallicity ([Fe/H]~$\sim$~--1.15~dex). If, 
however, we use the spectroscopically determined [Fe/H] values 
instead of the photometrically determined ones in the distance 
estimation of \citet{ive08}, we find a nearly constant systematic 
difference of $\sim$10\% for the entire metallicity range shown 
in Figure \ref{fig:distdiff}.

Using the distance, radial velocity, and proper motion of each 
star, the full space motions (U,V,W) were determined using the 
method in \citet{joh87} with an updated galactic coordinate 
transformation. We used a modified version of the \textit{gal\_uvw} 
routine from the IDL Astronomy User's 
Library\footnote{http://idlastro.gsfc.nasa.gov/} 
and the solar motion values from \citet{sch10} to correct to the
local standard of rest. The uncertainties in each of the velocity
components are on average $\sim$20~km~s$^{-1}$ for our sample
of stars.

\section{Analysis\label{sec:ana}}
The cuts described in the previous section were applied to provide
a sample of dwarf stars that could be used to isolate likely members
of the disk component(s) of our Galaxy. As was described in Section 
\ref{sec:intro}, past studies have used various methods to select and 
characterize the disk. We restrict our samples to 7.0~$<R<$~10.5 kpc 
since these radial distances provide us with the largest statistics and 
to avoid possible contamination from the bulge of the Galaxy.  In 
the vertical direction we limit the sample to 0.5~$<|Z|<$~3.0 kpc 
for determining radial metallicity gradients since there are few stars 
below these heights and the statistics fall rapidly above them. For 
vertical metallicity gradients we fit in the range 1.0~$<|Z|<$~3.0 
kpc to be more consistent with previous works and for reasons that 
will be discussed in Section \ref{sec:res}. These restrictions in 
radial and vertical distances are shown in Figure \ref{fig:spatial} 
with solid lines.

To determine the extent to which the thin disk contributes to our 
dwarf star sample we use the kinematic 
properties of the different components of the Galaxy as 
a discriminant. To do this, we used Equation 1 of \citet{ben03} and 
the characteristic velocity dispersions and asymmetric drift values 
given in their Table 1 to determine the likelihood for each star of 
belonging to either the thin disk, thick disk, or halo based on its 
kinematics alone. The likelihood for each star of belonging to a 
particular component of the Galaxy also depends on the local number 
density expected of that component. Since \citet{ben03} use local 
densities for the solar neighborhood and our sample of stars is at 
higher vertical heights, we use the Galactic model given by 
\citet{jur08} in column 2 of their Table 10. This is their best-fit 
model determined directly from the number density distribution maps 
found in their work.  Combining the kinematic and local number density 
likelihoods, for each individual star we determine the likelihood of 
it belonging to the thin disk, thick disk, or halo. Ratios of these 
likelihoods were then used to find stars that are more likely to be 
thin disk than thick disk (i.e. TD/D~$<$~0.10) and more likely to be 
thick disk than thin disk or halo (i.e. TD/D~$>$~10 and TD/H~$>$~10). 
These limits of being ten times more likely is in agreement with other 
papers employing the same technique \citep[e.g.][]{cos12}. This helps 
separate stars with kinematics most obviously matching either the thin 
or thick disks at the heights in which both populations overlap. We 
note that systematics in our distance determinations can affect the 
local number densities for each star and can also lead to incorrect 
velocity components, and therefore incorrect likelihoods for 
belonging to one Galactic component or another.  This possible issue 
is mitigated by the fact that we use strong cuts on the populations 
(i.e. being ten times more likely to be one than the other) and only 
use the likelihoods to separate our sample at the smallest vertical 
heights.

\begin{figure}[tb]
\plotone{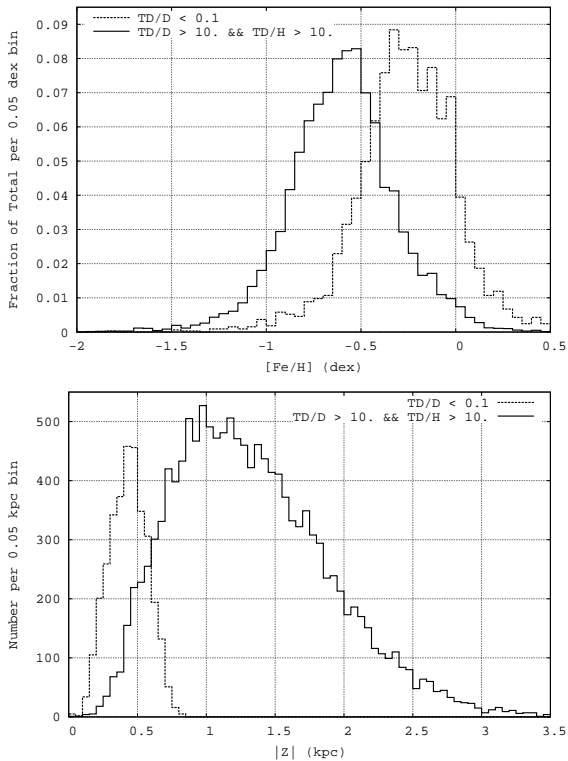}
\caption{Distribution in $|Z|$ (lower) and [Fe/H] (upper) for stars 
selected according to probabilities from spatial and kinematic 
information as described in the text.\label{fig:kinsep}}
\end{figure}

In the lower panel of Figure \ref{fig:kinsep} we show that the 
vertical distributions for stars defined as likely thin disk 
and likely thick disk are very different. Thin disk stars are peaked at 
heights of slightly less than 0.5~kpc and have no stars with heights 
larger than $\sim$0.75~kpc. Meanwhile, the thick disk distribution is 
much broader and has a peak slightly higher than 1~kpc. The upper 
plot in Figure \ref{fig:kinsep} shows that the kinematically selected 
thick disk sample has a peak metallicity near --0.6~dex, while the thin 
disk sample is peaked at a higher metallicity (near --0.2~dex). Both 
the vertical height distribution and the mean metallicities of the 
kinematically selected components are consistent with previous works. 
Of note is the fact that it is possible to separate the thin and 
thick disks in SDSS data using their kinematics and distances. Also, 
we see that for vertical heights of 0.5~$<|Z|<$~1.0 kpc our dwarf 
star sample contains both thin and thick disk stars ($\sim$80\% of 
the kinematically selected disk stars are thick disk), while above 
$|Z|$~=~1 kpc we only have thick disk stars.

Using the large sample of dwarf stars with reliably determined 
properties and a number of previous values with which to compare, 
we can now provide a better view of the overall abundance patterns 
of dwarf stars in the Galactic thick disk in both the radial and 
vertical directions.

\section{Results\label{sec:res}}
When looking for metallicity gradients of the disk many factors are
involved that could affect the outcome. The population of stars, their
locations in the Galaxy, and their kinematics are all distinguishing
characteristics that can (and do) give different results. In this
paper we use exclusively dwarf stars that are mostly from the 
thick disk component of our Galaxy. We have a very large sample,
however, with a wide range of distances and kinematics. As such, we
divide our sample based on these properties. Specifically, as discussed
in the previous section, we find that for vertical heights of 
$|Z|>$~1 kpc our sample has no thin disk contamination and for 
0.5~$<|Z|<$~1.0 kpc the contamination is approximately 20\%.
Furthermore, we group our stars by their positions 
in order to disentangle the affect of one metallicity gradient on
the other. For example, if we simply took our entire sample
and fit for a vertical metallicity gradient using all radial distances,
the contribution of a radial metallicity gradient could have an affect
on the vertical metallicity gradient. We choose bin sizes
of 0.5~kpc in order to have a statistically significant sample of
stars in each bin and minimize the affect of the other gradient. We 
also provide metallicity gradients without binning to see how these 
values agree with those previously published.

It is important to note that while we do have stars in the full range
of fitted values, not all the bins will have the same distribution
of radial and/or vertical distances. This could affect our 
gradients. By averaging the points in each bin, however, and then 
determining the gradients from these averaged values we reduce the 
affect of these changes and any potential outliers in our sample. 

To estimate errors in the metallicity gradients, a bootstrapping method
was employed where we randomly re-sampled half the number of stars
in each group one thousand times and fit each one for a gradient.
The errors quoted in the tables that follow are the standard deviations
of the results from this method.  To determine the affect of a 
systematic difference in the distance estimates we have included 
metallicity gradients using both the isochrone 
and photometric distance methods.

\subsection{Radial Gradients\label{ssec:radgrad}}
We first look for metallicity gradients in the radial direction for
our sample stars. The presence of a radial metallicity gradient in the 
Galactic plane has been observed using various methods (blue stars, open 
clusters, nebulae, etc). A radial metallicity gradient away from the 
Galactic plane is not as well established. In fact, depending on the 
population and height above the plane, different results are found. 
\citet{all06} find no evidence of a radial metallicity gradient for 
stars in the range 1~$<|Z|<$~3 kpc
and while \citet{nor04} find a positive gradient for their old age
sample ($>$10~Gyr), their value is still consistent with no gradient
within errors. Recent results from RAVE have found positive gradients
for likely thick disk stars selected in various ways and their results
are consistent with a positive, nonzero gradient within errors. Radial
metallicity gradients from various references can be found in Table
\ref{tab:rad}.

\begin{table*}[tb]
\begin{center}
\caption{Radial metallicity gradients from the literature.\label{tab:rad}}
\begin{tabular}{ c c c r }
\hline
 & \textbf{gradient }(dex kpc$^{-1}$) & \textbf{error} & \textbf{notes} \\
\hline
\citet{all06} & $\sim$0 & & 1$<|Z|<$3 kpc \\
\hline
\citet{nor04} & +0.028 & 0.036 & age $>$10 Gyr \\
\hline
\multirow{4}{*}{\citet{cos12}} & +0.016 & 0.011 & F dwarfs, TD/D$>$10 \\
 & +0.016 & 0.012 & F dwarfs, e$_v>$0.1 \\
 & +0.010 & 0.009 & G dwarfs, TD/D$>$10 \\
 & +0.037 & 0.016 & G dwarfs, e$_v>$0.1 \\
\hline
\citet{ruc11} & +0.01 & 0.04 & [Fe/H]$<$--1.2 \\
\hline
\multirow{2}{*}{\citet{cheng12}}
 & +0.0028 & $^{+0.0071}_{-0.0052}$ & 1$<|Z|<$1.5 kpc \\
 & --0.013 & $^{+0.0093}_{-0.0016}$ & 0.5$<|Z|<$1.0 kpc \\
\hline
\end{tabular}
\end{center}
\end{table*}

\begin{figure}[tb]
\plotone{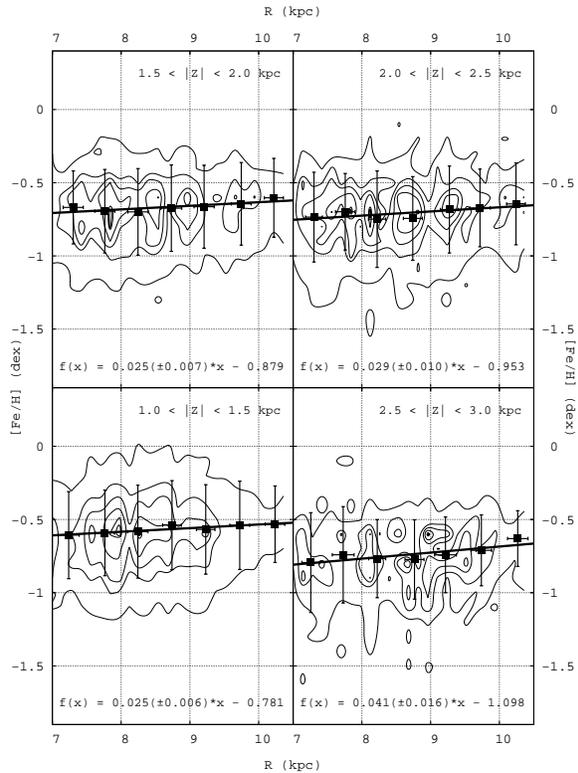}
\caption{Fits for the radial metallicity gradient of the dwarf star 
sample in different $|Z|$ ranges.  Contour lines are shown for 
95\%, 90\%, 68\%, 50\%, 33\%, and 10\% of the peak density.
\label{fig:grIndR}}
\end{figure}

\begin{figure}[tb]
\plotone{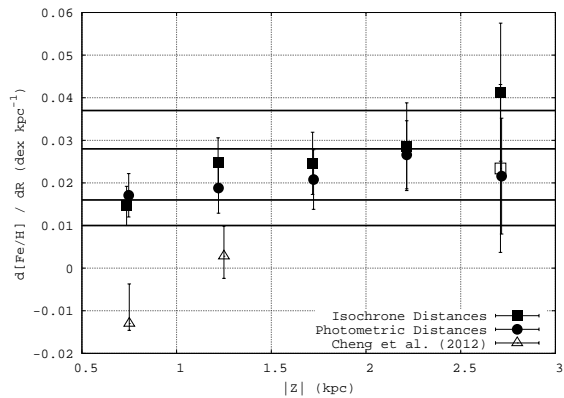}
\caption{Radial metallicity gradients of our dwarf star sample found 
using distances from the isochrone (solid squares) and photometric (solid 
circles) methods for different heights above the Galactic plane. The 
horizontal lines and open triangles correspond to previously published 
values (see Table \ref{tab:rad}).\label{fig:grdsR}}
\end{figure}

Since we have such a large sample, we are able to look not only for
a radial metallicity gradient, but also see how the gradient changes
as a function of height above the Galactic plane. In Figure 
\ref{fig:grIndR} we show metallicity as a function of the
radial distance from the Galactic center for our entire 
sample of stars for different ranges in the vertical height. 
The resulting radial metallicity gradients can be found in Table 
\ref{tab:ourR} and a plot of the radial metallicity gradient as a 
function of average $|Z|$ is shown in Figure \ref{fig:grdsR}. 
For reference, horizontal lines have been drawn in Figure \ref{fig:grdsR} 
for the results of previous publications found in Table \ref{tab:rad}
and two overlapping points in $|Z|$ from \citet{cheng12} have been
included as well. For vertical heights of 1~$<|Z|<$~3 kpc we find that 
the radial metallicity gradients are relatively constant as a function 
of $|Z|$. In the vertical height bin where thin disk stars contaminate 
our sample (0.5~$<|Z|<$~1.0 kpc) the radial metallicity gradient is 
slightly lower. In the highest vertical height bin (2.5~$<|Z|<$~3.0 kpc) 
the number of stars is an order of magnitude smaller than the bins with 
lower vertical heights and the difference between the results using the 
two different distance estimates is largest. Furthermore, from 
Figure \ref{fig:grIndR} one can see that in this vertical height range 
the point for radial distances of 10.0~$<R<$~10.5 kpc using the isochrone 
method is noticeably above the fit and forces the linear fit to be 
steeper. Disregarding this point when fitting for the metallicity 
gradient gives a value (+0.023 dex kpc$^{-1}$) that is more consistent 
with lower heights and the photometric distance estimate. We have 
included this value in Figure \ref{fig:grdsR} as the open square.

We could fit for a radial metallicity gradient for the entire range 
associated with the thick disk (1~$<|Z|<$~3 kpc), however, as is evident 
from the offsets in the linear fits shown in Figure \ref{fig:grIndR}, 
there is a strong vertical metallicity gradient in the sample that will 
affect an overall determination in the radial direction.

\begin{table*}[tb]
\begin{center}
\caption{Radial metallicity gradients of our dwarf star sample.
\label{tab:ourR}}
\begin{tabular}{ c r@{,}l c c c r@{,}l c c c }
\hline
 & \multicolumn{5}{c}{\textbf{Isochrone Distance Method}}
 & \multicolumn{5}{c}{\textbf{Photometric Distance Method}} \\
\textbf{$|Z|$ Height} 
 & \multicolumn{2}{c}{$N$} & $\overline{|Z|}$
 & $\overline{R}$ & \textbf{Gradient}
 & \multicolumn{2}{c}{$N$} & $\overline{|Z|}$
 & $\overline{R}$ & \textbf{Gradient} \\
(kpc)
 & \multicolumn{2}{c}{stars} & (kpc) & (kpc) & (dex kpc$^{-1}$)
 & \multicolumn{2}{c}{stars} & (kpc) & (kpc) & (dex kpc$^{-1}$) \\
\hline
0.5--1.0
 & 16&892 & 0.74 & 8.46 & +0.015$\pm$0.005
 & 15&150 & 0.75 & 8.52 & +0.017$\pm$0.005 \\
\hline
1.0--1.5
 &  8&881 & 1.22 & 8.47 & +0.025$\pm$0.006
 &  9&424 & 1.22 & 8.50 & +0.019$\pm$0.006 \\
\hline
1.5--2.0
 &  4&314 & 1.72 & 8.53 & +0.025$\pm$0.007
 &  4&954 & 1.72 & 8.56 & +0.021$\pm$0.007 \\
\hline
2.0--2.5
 &  1&907 & 2.22 & 8.59 & +0.029$\pm$0.010
 &  2&519 & 2.22 & 8.56 & +0.027$\pm$0.008 \\
\hline
2.5--3.0
 & \multicolumn{2}{r}{730} & 2.71 & 8.68 & +0.041$\pm$0.016\tablenotemark{a}
 &  1&136 & 2.72 & 8.64 & +0.022$\pm$0.014 \\
\hline
\end{tabular}
\tablenotetext{a}{Excluding the point with radial distances of 
10.0~$<R<$~10.5 kpc gives a metallicity gradient of +0.023 dex kpc$^{-1}$.}
\end{center}
\end{table*}

Radial metallicity gradients determined using the photometric method 
distances are very similar to those found with the isochrone method. 
Indeed, the values all agree with each other within errors. The fact 
that there is no large discrepancy between the results using the two 
methods means that systematic differences in the distances on the 
order we observe between the two distance estimates (up to $\sim$20\%) 
do not significantly affect our results.

Our radial metallicity gradients lie in the middle of the previously 
published values and agree most closely with those of \citet{nor04} 
and the F dwarf samples of \citet{cos12}. Our results differ from 
those of \citet{cheng12} in the bins that overlap with their work. 
As is shown in Figure \ref{fig:loggVabun}, there is a noticeable 
change in the metallicity distribution for stars with surface 
gravities below $\sim$4.4~dex in our sample, which could be the 
reason for the discrepancy between our work using high surface 
gravity dwarfs and their work using turnoff stars with lower surface 
gravities. It is also possible that there are different ratios of 
thin and thick disk stars between our sample and theirs, which would 
obviously give a different metallicity gradient if the two 
populations themselves had gradients different from each other.

\subsection{Vertical Gradients\label{ssec:vertgrad}}
In a similar fashion to what was done for the radial gradients, we
examine the vertical metallicity gradients of our sample. In Table
\ref{tab:vert} we list some previously published values using various
populations and sample selection methods.

\begin{table*}[tb]
\begin{center}
\caption{Vertical metallicity gradients from the literature.\label{tab:vert}}
\begin{tabular}{ c c c r }
\hline
 & \textbf{gradient} (dex kpc$^{-1}$) & \textbf{error} & \textbf{notes} \\
\hline
\citet{all06} & $\sim$0 &  & 1~$<|Z|<$~3 kpc \\
\hline
\multirow{2}{*}{\citet{chen11}} & --0.120 & 0.01 & Galaxy model and fits \\
 & --0.225 & 0.07 & All points \\
\hline
\citet{katz11} & --0.068 & 0.009 & 2 lines of sight \\
\hline
\citet{ruc11} & --0.09 & 0.05 & {[}Fe/H{]}$<$--1.2 \\
\hline
\citet{kor11} & --0.14 & 0.05 & 1~$<|Z|<$~4 kpc \\
\hline
\end{tabular}
\end{center}
\end{table*}

We bin in radial distance to fit for vertical metallicity gradients 
and restrict our fits to vertical heights of 1~$<|Z|<$~3 kpc for two 
reasons. The first is to be more consistent with previously published 
values. The second is to provide a clearer picture of the vertical 
metallicity gradient of only the thick disk. There is a significant 
overlap of the thin and thick disks for vertical heights of 
$|Z|<$~1~kpc, which can be seen in the vertical distribution of the 
kinematically separated samples shown in Figure \ref{fig:kinsep}. 
Therefore, if we extend our fits below this height for our full 
sample of stars we will not be probing a single population of our 
Galaxy, but the contribution from both the thin and thick disks. As 
we will show, the kinematically selected thick disk with $|Z|<$~1~kpc 
could be included in the fits without significantly altering the 
results, but in order to be more consistent with previous results 
we use only vertical heights of 1~$<|Z|<$~3 kpc.

\begin{figure}[tb]
\plotone{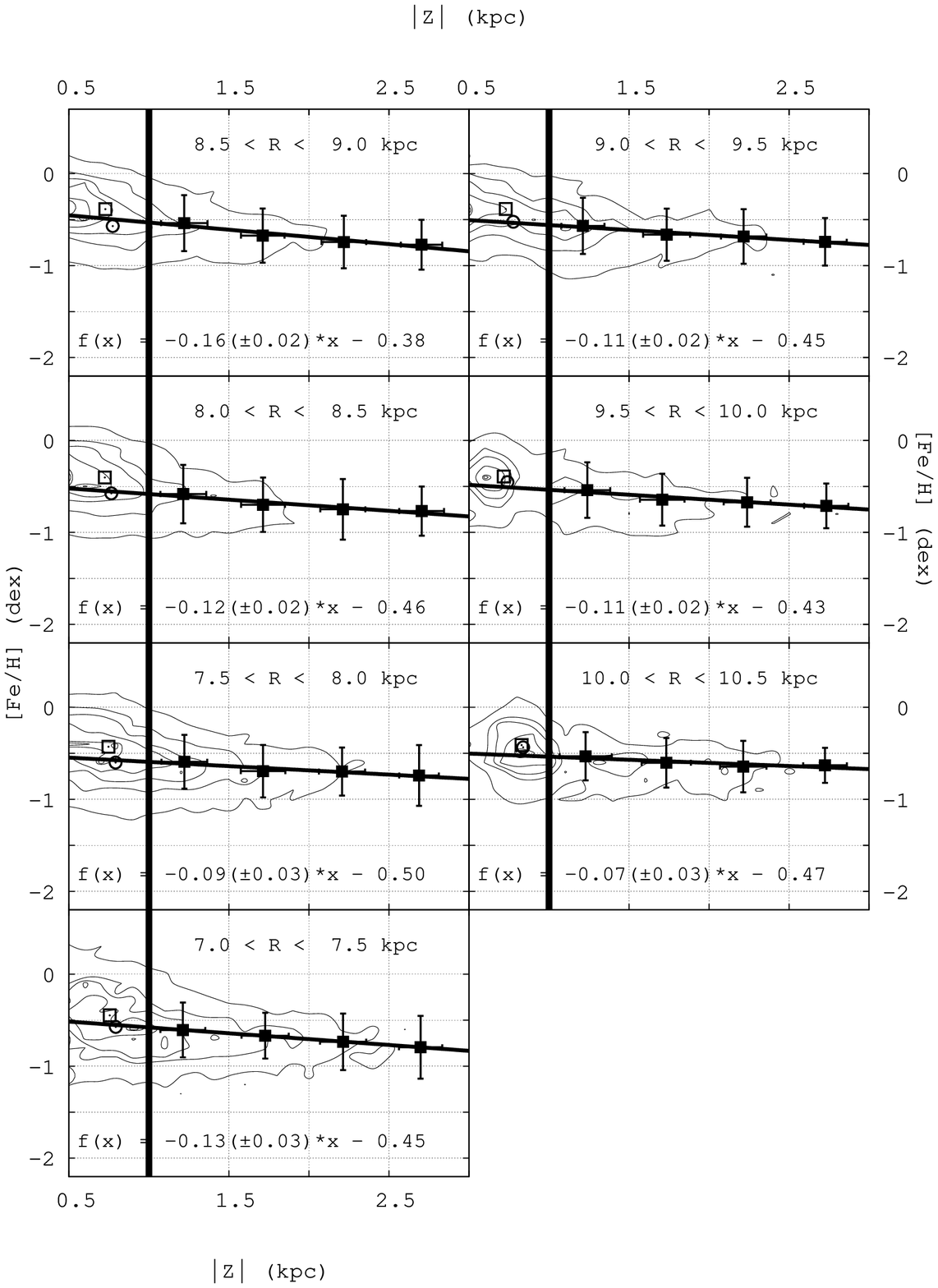}
\caption{Fits for the vertical metallicity gradient of the dwarf star 
sample in different $R$ ranges. Contour lines are shown for 
95\%, 90\%, 68\%, 50\%, 33\%, and 10\% of the peak density.
\label{fig:grIndZ}}
\end{figure}

Vertical metallicity gradients for our sample of stars are given in 
Table \ref{tab:ourV}. Plots showing the vertical metallicity gradients 
in different radial bins can be found in Figure \ref{fig:grIndZ}. 
This plot shows the effect of selecting only stars with heights above 
$|Z|$~=~1~kpc. The points in the 0.5~$<|Z|<$~1.0 kpc bin show the 
average metallicity of the full sample (open squares) and of the 
kinematically selected thick disk sample (open circles). Careful 
examination of these points shows that the full sample has a 
higher metallicity and is systematically above the fit value. 
This is most certainly due to the contribution of the more 
metal-rich thin disk in our sample at these heights and is the reason 
we have excluded this region from our metallicity gradient determinations. 
The kinematically selected thick disk sample matches well with the fit 
value.

\begin{figure}[tb]
\plotone{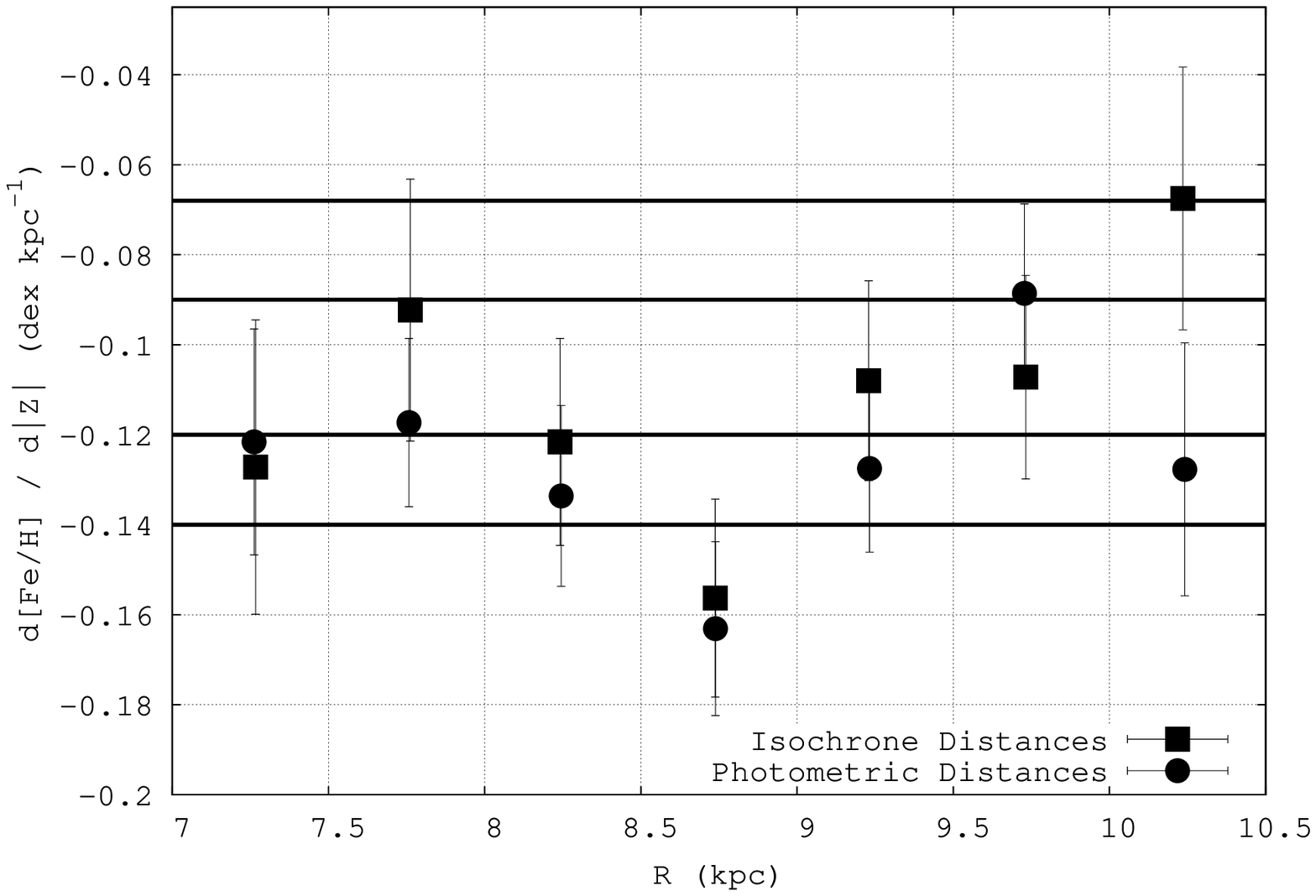}
\caption{Vertical metallicity gradients of our dwarf star sample found 
using distances from the isochrone (solid squares) and photometric (solid 
circles) methods for different radial distances from the Galactic center.  
The horizontal lines correspond to previously published values (see Table 
\ref{tab:vert}).\label{fig:grdsZ}}
\end{figure}

A plot of the vertical metallicity gradients as a function of radial 
distance from the Galactic center is found in Figure \ref{fig:grdsZ}. 
It is obvious that the metallicity gradient for the largest radial 
distances are the most discrepant between the two distance 
determinations. As was noted for the radial metallicity gradients, 
these distances have the lowest number of stars making statistics 
a possible issue in the metallicity gradient determination. From Figure 
\ref{fig:grIndZ} one can see that including the 0.5~$<|Z|<$~1.0 kpc 
bin in the fit for radial distances of 10.0~$<R<$~10.5 kpc found 
using the isochrone method would give a steeper slope. This would bring 
the result closer to the values at smaller Galactic radii as well as 
the result using photometric distances. As was stated 
above, however, for consistency with previous results we have 
restricted our fits to vertical heights of 1~$<|Z|<$~3 kpc.

\begin{table*}[tb]
\begin{center}
\caption{Vertical metallicity gradients of our dwarf star sample 
with heights of 1~$<|Z|<$~3~kpc.\label{tab:ourV}}
\begin{tabular}{ r@{--}l r@{,}l c c c r@{,}l c c c }
\hline
\multicolumn{2}{c}{}
 & \multicolumn{5}{c}{\textbf{Isochrone Distance Method}}
 & \multicolumn{5}{c}{\textbf{Photometric Distance Method}} \\
\multicolumn{2}{c}{\textbf{$R$ Distance} }
 & \multicolumn{2}{c}{$N$} & $\overline{|Z|}$
 & $\overline{R}$ & \textbf{Gradient}
 & \multicolumn{2}{c}{$N$} & $\overline{|Z|}$
 & $\overline{R}$ & \textbf{Gradient} \\
\multicolumn{2}{c}{(kpc)}
 & \multicolumn{2}{c}{stars} & (kpc) & (kpc) & (dex kpc$^{-1}$)
 & \multicolumn{2}{c}{stars} & (kpc) & (kpc) & (dex kpc$^{-1}$) \\
\hline
7.0&7.5
  & 1&812 & 1.58 & 7.27 & --0.13$\pm$0.03
  & 2&042 & 1.67 & 7.26 & --0.12$\pm$0.03 \\
\hline
7.5&8.0
  & 3&125 & 1.52 & 7.76 & --0.09$\pm$0.03
  & 3&473 & 1.57 & 7.76 & --0.12$\pm$0.02 \\
\hline
8.0&8.5
  & 3&427 & 1.48 & 8.24 & --0.12$\pm$0.02
  & 3&827 & 1.52 & 8.25 & --0.13$\pm$0.02 \\
\hline
8.5&9.0
  & 2&913 & 1.54 & 8.74 & --0.16$\pm$0.02
  & 3&286 & 1.57 & 8.74 & --0.16$\pm$0.02 \\
\hline
9.0&9.5
  & 2&296 & 1.55 & 9.23 & --0.11$\pm$0.02
  & 2&674 & 1.59 & 9.23 & --0.13$\pm$0.02 \\
\hline
9.5&10.0
  & 1&423 & 1.65 & 9.73 & --0.11$\pm$0.02
  & 1&684 & 1.67 & 9.73 & --0.09$\pm$0.02 \\
\hline
10.0&10.5
  & \multicolumn{2}{r}{836} & 1.67 & 10.24 & --0.07$\pm$0.03
  & 1&047 & 1.74 & 10.24 & --0.13$\pm$0.03 \\
\hline
\hline
7.0&10.5
  & 15&832 & 1.54 & 8.51 & --0.113$\pm$0.010
  & 18&033 & 1.59 & 8.53 & --0.125$\pm$0.008 \\
\hline
\end{tabular}
\end{center}
\end{table*}

There is some structure to the gradients as a function of 
radial distance but no obvious trends.  In fact, a single 
value can describe our results. Because the radial 
metallicity gradient is much smaller (by an order of magnitude) 
than the vertical metallicity gradient, we can fit the data 
for stars with radial distances of 7.0~$<R<$~10.5 kpc and 
vertical heights of 1~$<|Z|<$~3~kpc, which gives us values of 
--0.113~$\pm$~0.010 and --0.125~$\pm$~0.008 dex~kpc$^{-1}$ for 
our full dwarf star sample using isochrone and photometric distances 
respectively.  These values are in very good agreement with the 
shallower vertical metallicity gradient found by \citet{chen11} 
as well as those given in \citet{kor11} and \citet{ruc11}. It 
is slightly steeper than the result of \citet{katz11}.

Vertical metallicity gradients found using the photometric 
distance determination are very similar to those found using 
the isochrone distance determination. The values from each 
method agree within errors and there is no obvious systematic 
difference between the two. The only discrepancy between 
the metallicity gradients determined using the different distance 
estimates is for the bin with 10.0~$<R<$~10.5 kpc, which also has 
the lowest statistics.

\section{Discussion\label{sec:disc}}
We have probed the metallicity gradients in both the radial and 
vertical directions using dwarf stars selected from SDSS. We 
kinematically separate this sample into thin and thick disk 
components at vertical heights where the two populations are 
known to overlap using probabilities defined and used in previous 
analyses. This is significant because this separation is most 
often applied to stars in the solar neighborhood, but we have 
shown that it can be successfully used for SDSS data with vertical 
heights of up to 1~kpc by taking into account the local number density 
as a function of height.

Stars associated with the Galactic thick disk have a positive 
radial metallicity gradient for heights of 1~$<|Z|<$~3 kpc. 
Our results seem to be at odds with the results of \citet{cheng12} in 
the two vertical height bins that overlap. However, the different 
populations and selection methods between our work and theirs make 
it difficult to compare the two results directly. Our results 
indicate that a radial metallicity gradient determined for stars 
with $|Z|<$~1.0 kpc will heavily depend on the 
fraction of thin and thick disk stars in the sample if the 
two populations have very distinct values and trends.

When fitting the entire sample of stars with vertical distances 
most consistent with the thick disk, we find a radial metallicity
gradient that is positive and not consistent with zero within errors.
In particular, we find radial metallicity gradients of 
+0.025 $\pm$~0.006 (+0.019 $\pm$~0.006), 
+0.025 $\pm$~0.007 (+0.021 $\pm$~0.007), 
+0.029 $\pm$~0.010 (+0.027 $\pm$~0.008), and 
+0.041 $\pm$~0.016 (+0.022 $\pm$~0.014) dex~kpc$^{-1}$ 
using isochrone (photometric) distances for stars in 0.5~kpc bins 
in vertical height between 1~$<|Z|<$~3 kpc and with radial distances 
of 7.0~$<R<$~10.5 kpc. Restricting the fit for the highest 
vertical height bin to 7.0~$<R<$~10.0 kpc changes the result using 
isochrone distances from +0.041 to +0.023 dex~kpc$^{-1}$, which agrees 
better with lower heights and with the photometric distance 
determination. To our knowledge, this is the first time that a 
radial metallicity gradient for the thick disk has been found at these 
vertical heights. The values we find are in very good agreement with the 
old stellar sample value of \citet{nor04} and with both of the thick disk 
F dwarf samples of \citet{cos12}. It also agrees within errors with the 
results of \citet{ruc11}. Due to the size of our sample, our statistical 
error is smaller than these previous works.

Fitting the full sample of stars with radial distances of 7.0~$<R<$~10.5 
kpc in the range 1~$<|Z|<$~3 kpc, we find a vertical metallicity gradient 
of --0.113~$\pm$~0.010 (--0.125~$\pm$~0.008) dex~kpc$^{-1}$ using 
isochrone (photometric) distances. This value is in good agreement 
with the results of \citet{chen11}, \citet{kor11}, and \citet{ruc11}.

The results of \citet{bovy12} (among others) suggests that the Galactic 
disk is a single component that changes smoothly as a function of both 
radial and vertical distances. We observe no obvious or significant 
trends in either the radial or vertical metallicity gradients for 
stars most consistent with the thick disk of our Galaxy.  However, 
it is not clear if the kinematic selection, distances, and/or bin 
sizes we have used would be able to detect changes in the metallicity 
gradient due to such a scenario. Furthermore, as discussed in their 
work, dividing the disk into components using different methods can 
give conflicting results, making it difficult to compare the results 
from geometric, kinematic, and abundance decompositions.

In summary, the metallicity gradients we find in both the radial and 
vertical directions using the sample of dwarf stars selected from SDSS 
DR8 show no large or obvious trends in the distance ranges we are 
able to probe. The vertical metallicity gradient we have found is much 
steeper than the radial metallicity gradient, but we have found both to 
be nonzero. Future large spectroscopic survey programs, such as LAMOST 
\citep{lamost}, should provide greater statistics and more complete 
samples of stars in a wider range of distances so that metallicity 
gradients can be probed in more detail and an even better 
understanding of our Galactic disk can be realized.

\acknowledgments

We would like to thank the anonymous referee for their helpful 
comments and suggestions for improving the paper. 
We would also like to thank Dr. Martin Smith and Dr. Ron Wilhelm for 
their comments and suggestions.

Funding was provided by the National Natural Science Foundation of 
China under grants 11150110135, 11073026 and 11222326 and the Chinese 
Academy of Sciences under grant KJCX2-YW-T22 and the fellowship for 
young international scientists.

Funding for SDSS-III has been provided by the Alfred P. Sloan Foundation,
the Participating Institutions, the National Science Foundation, and
the U.S. Department of Energy Office of Science. The SDSS-III web
site is http://www.sdss3.org/.

SDSS-III is managed by the Astrophysical Research Consortium for the
Participating Institutions of the SDSS-III Collaboration including
the University of Arizona, the Brazilian Participation Group, Brookhaven
National Laboratory, University of Cambridge, University of Florida,
the French Participation Group, the German Participation Group, the
Instituto de Astrofisica de Canarias, the Michigan State/Notre Dame/JINA
Participation Group, Johns Hopkins University, Lawrence Berkeley National
Laboratory, Max Planck Institute for Astrophysics, New Mexico State
University, New York University, Ohio State University, Pennsylvania
State University, University of Portsmouth, Princeton University,
the Spanish Participation Group, University of Tokyo, University of
Utah, Vanderbilt University, University of Virginia, University of
Washington, and Yale University.

\end{document}